\documentclass[12pt]{article} 
 
\usepackage{epsfig} 
\usepackage{amssymb} 
\usepackage{amsmath} 
\usepackage{amsfonts} 
\usepackage{cite}

\newcommand{\be}{\begin{equation}} 
\newcommand{\beq}{\begin{equation}} 
\newcommand{\ee}{\end{equation}}

\newcommand{\bea}{\begin{eqnarray}} 
\newcommand{\eea}{\end{eqnarray}} 
 
\newcommand{\eq}[1]{(\ref{#1})} 
 
\newcommand{\as}{\bar\alpha_s}

\newcommand{\wid}{0.7\columnwidth}
\DeclareMathOperator{\lid}{Li_2}
\DeclareMathOperator{\lit}{Li_3}

\begin{document} 

\begin{flushright} 
CPHT--RR 048.0805 \\
LBNL--58953\\ 
hep-ph/0508134 
\end{flushright} 
 
\vspace{\baselineskip}

\begin{center} 
 
\textbf{\LARGE BFKL resummation effects in   
 $\gamma^*\gamma^*\to \rho \rho$}\\ 
\vspace{3\baselineskip}

{\large 
R.\ Enberg$^{a,b}$, 
B.\ Pire$^a$, 
L.\ Szymanowski$^{c,d}$ and 
S.\ Wallon$^e$ 
}\\

\vspace{2\baselineskip}

${}^a$\,CPHT\footnote{Unit{\'e} mixte C7644 du CNRS}, 
\'Ecole Polytechnique, 91128 Palaiseau, France \\[0.5\baselineskip]  
${}^b$\,Lawrence Berkeley 
National Laboratory, Berkeley, CA 94720, USA\\[0.5\baselineskip]  
${}^c$\,So{\l}tan Institute for Nuclear Studies, Ho\.za 69, 00-681 Warsaw, Poland\\[0.5\baselineskip]
${}^d$\,Universit\'e  de Li\`ege,  B4000  Li\`ege, Belgium\\[0.5\baselineskip] 
${}^e$\,LPT\footnote{Unit{\'e} mixte 8627 du CNRS}, Universit\'e 
Paris-Sud, 91405-Orsay, France \\

\vspace{5\baselineskip} 
\textbf{Abstract}\\ 
\vspace{1\baselineskip} 
\parbox{0.9\textwidth}{We calculate the leading order BFKL amplitude  
 for the exclusive diffractive process $\gamma^*_L (Q_1^2)\gamma^*_L(Q_2^2) \to 
\rho_L^0 \rho_L^0$ in the forward direction, 
which can be studied in future high energy  $e^+e^-$ linear colliders.  
The resummation effects are very large compared to the fixed-order calculation. We also estimate the next-to-leading logarithmic corrections to the amplitude by using a specific resummation of higher order effects and find a substantial growth with energy, but smaller than in the leading logarithmic approximation. 
}

\end{center}

\setcounter{footnote}{0}

\newpage

\section{Introduction} 
 
\begin{figure}[tbp]
\begin{center}
\epsfig{file=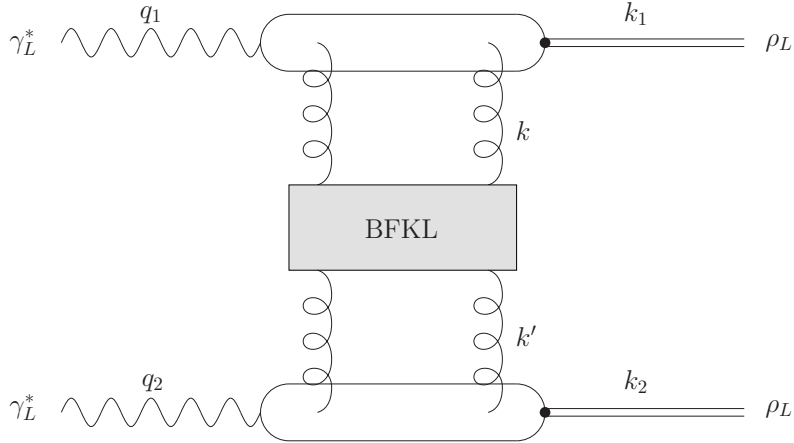,width=\wid}
\end{center}
\caption{The process 
$\gamma_L^*(q_1)\;\gamma_L^*(q_2) \to \rho^0_L(k_1) \;\rho^0_L(k_2)$. 
The box represents the BFKL Green's function, and the $t$-channel gluons couple to the quark lines in all possible ways.
\label{diagram}}
\end{figure}

The next generation of $e^+e^-$ colliders will offer the possibility of 
clean testing of QCD dynamics.  
 By selecting events in which two vector mesons are produced 
with a large rapidity gap in between, through the scattering of two highly virtual photons, 
one accesses a kinematical regime in which a perturbative 
approach is justified. If additionally one selects events with 
comparable photon virtualities, the perturbative Regge dynamics of QCD 
 should dominate and allow the use of resummation techniques  of the BFKL \cite{bfkl} type. 
In this paper we  study these effects in the case of the reaction
(see Fig.\ \ref{diagram})
 \beq 
\label{process} 
\gamma_L^*(q_1)\;\gamma_L^*(q_2) \to \rho_L(k_1)  \;\rho_L(k_2)\, 
\ee 
in the forward region, i.e.\  $-t=-(q_1-k_1)^2=-t_{min}$, 
where the virtualities  $Q_1^2 =-q_{1}^2$, $Q_2^2=-q_{2}^2$ 
 of the scattered photons play the role of the hard  scales.

This is an interesting process to study, first because the experimental signature is simple:\ double tagged events in high energy $e^+ e^-$ collisions \cite{budnev}, with two longitudinally polarized rho mesons in the final state. The non-zero virtualities of the photons allow studying also the forward region. Second, the theoretical calculation can be performed analytically and gives a simple answer which makes it easy to investigate BFKL effects. This is partly because in the forward region we are studying, the amplitude for transversely polarized photons vanishes identically \cite{PSW}.

No data exist yet for the process \eq{process} at sufficiently high energies,
but some experimental data exist  for $Q_2^2$ small  \cite{L3} which 
may be analyzed \cite{apt} in terms of generalized distribution amplitudes 
\cite{gda}. 
Related reactions have been calculated, both at the Born level and using BFKL resummation. The similar process with real incoming photons, $\gamma \gamma \to V_1 V_2$ and large $t$ where the hard scale is provided by the momentum transfer instead of the photon virtualities, was studied in \cite{Ginzburg:1985tp,2jpsi},  
and the process $\gamma p \to V X$ at large $t$, has been studied both for heavy \cite{HeavyVM} and light \cite{LightVM} vector mesons. The advantage of the latter process is that there now exists high quality data from HERA \cite{HERA}, and the BFKL calculations reproduce the measured differential cross sections for longitudinal $\rho$ meson production and for $J/\psi$ production very well, while the Born level calculations do not work so well \cite{HeavyVM,LightVM,HERA}. 
At the same time the cross section for transversely polarized $\rho$ mesons is not well understood \cite{LightVM}.

In a recent paper \cite{PSW},
  the scattering amplitude of the process (\ref{process}) was calculated 
  in the Born 
approximation and  the feasibility of a dedicated experiment was proved. The 
production amplitude decreases dramatically with $t$ so that its magnitude at 
$t=t_{min}$  
dictates the rate of the reaction. In this 
paper, we report on a calculation of BFKL effects in the leading order 
approximation 
at $t=t_{min}$. We find that the leading logarithmic (LL)
BFKL resummation greatly enhances the cross section, as previously observed in 
other reactions \cite{bfklinc,Boonekamp,brodsky,2jpsi}. 
In addition, we estimate next-to-leading logarithmic (NLL) effects by the use
of the BLM scale fixing prescription \cite{BLM} supplemented by the resummation
scheme of Refs.~\cite{Salam,CCS,Khoze}. 
We show that there is still a substantial increase in the cross section 
compared to fixed-order calculations, but smaller 
than the LL calculation. It would be interesting to compare our estimates with the ones based on the 
recently available NLL impact factor for the 
process $\gamma^*\to V$ of Ref.~\cite{NLLIF} and the NLL BFKL kernel \cite{NLLBFKL}.

\section{Leading order BFKL forward amplitude} 

\subsection{General expression} 
 
In the BFKL framework, the amplitude of the process \eq{process} can be expressed through the inverse Mellin transform with
respect to the squared center-of-mass energy $s$  as
\begin{align} 
A(s,t) = is \int\frac{d\omega}{2\pi i} e^{\omega Y} f_\omega(r^2)\;, 
\label{ampldef} 
\end{align} 
where $t-t_{min} \sim -r^2,$ ($r$ is considered as Euclidean, 
as is any two-dimensional vector in the following), and $Y$ is the rapidity variable, $Y=\ln (s/s_0)$. The minimum momentum transfer is here given by $t_{min}\sim Q_1^2 Q_2^2/s$~\cite{PSW}.
In the particular case where $r^2=0$, 
the BFKL Green's function 
can be easily obtained in
momentum space \cite{Lipatovreview}. In this case, the 
 impact representation for 
$f_\omega(0)$ reads
\begin{align} 
f_\omega(0) = \frac{1}{2}\int  
\frac{dk^2}{k^3}\frac{dk'^2}{k'^3}  
\Phi^{ab}_1(k) \Phi^{ab}_2(k') 
\int_{-\infty}^{\infty} d\nu\;   
\frac{1}{\omega - \omega(\nu)}\, 
\left(\frac{k^2}{k'^2}\right)^{i\nu} \;,
\label{bfklamp3} 
\end{align} 
where the integration over angles has been performed. The functions $\Phi^{ab}_i$ are the impact factors describing the coupling of the BFKL pomeron to vertex 1 or 2, and $a,b$ are color indices. The function
 $\omega(\nu)$
is the BFKL characteristic function which is defined by \cite{bfkl}
\beq
\label{defomega}
\omega(\nu)=\as \chi(\nu)\;,
\ee
with $\as \equiv \alpha_s N_c /\pi$ and 
\beq
\label{defch}
\chi(\nu)= 2 \psi(1)-\psi\left(\frac{1}{2} +i \nu
\right)-\psi\left(\frac{1}{2} -i \nu \right)\,, \quad 
\psi(x)=\Gamma'(x)/\Gamma(x).
\ee
The general solution for arbitrary values of $t$ is more involved \cite{Lipatov:1985uk}
and will not be discussed here. We only want to stress that it has a simple 
expression at the Born level,
\begin{align} 
\label{ABorn}
A = i s \int \frac{d^2k}{ k^2( r-k)^2} 
\Phi^{ab}_1(k,r-k) \Phi^{ab}_2(k,r- k), 
\end{align} 
which has recently been evaluated in Ref.~\cite{PSW}. The special case $r^2=0$ of
Eq.~(\ref{ABorn}) can be readily obtained from Eq.~(\ref{bfklamp3}). This 
relation will be studied in Section \ref{born}.
 The  impact factor  
 $\gamma^*_L \to \rho_L$ is given at $t=t_{min}$ by \cite{PSW} 
\be 
\Phi^{ab}_i(k, r = 0) = {\cal C}^{ab} Q_i  \int_0^1 d z \, z \bar z \, \phi(z)  
\left[ \frac{1}{m_i^2} - \frac{1}{k^2 + m_i^2}  \right] \;, 
\label{IF2}
\ee 
where $m_i^2=Q_i^2 z \bar z$ (with $\bar z = 1-z$),
$\phi(z)$ is the distribution amplitude of the meson which is here given by its asymptotic form $\phi(z)=6 z\bar z$, and\footnote{The first factor $1/(2\pi)^2$ differs from \protect\cite{PSW} and comes from the convention for impact factors.}  
\be 
{\cal C}^{ab} = \frac{1}{(2\pi)^2} 16 \pi^2 \alpha_s \frac{e}{\sqrt{2}} \frac{\delta^{a b}}{2 N_c} f_\rho \;, 
\label{Cab}
\ee  
where $f_\rho$ is the $\rho^0-$meson decay constant.

Inserting \eq{IF2} and \eq{Cab} in \eq{bfklamp3} we get 
\begin{align} 
&f_\omega(r^2=0) =  
4 \pi \alpha_s^2 \alpha_{em} \frac{N_c^2-1}{N_c^2} f_{\rho}^2 Q_1 Q_2
\int_0^1 d z_1 \, z_1 \bar z_1 \, \phi(z_1)
\int_0^1 d z_2 \, z_2 \bar z_2 \, \phi(z_2)  \label{bfklamp} \\ 
&\times 
\int_{-\infty}^{\infty} d\nu\;   
\frac{1}{\omega - \omega(\nu,n)}
\left[  
   \int \frac{dk^2}{k^3} 
   \left(\frac{1}{m^2} - \frac{1}{k^2 + m^2}\right)   
   {k}^{2i\nu} 
\right] 
\left[ 
   \int \frac{dk'^2}{k'^3}  
   \left(\frac{1}{m^2} - \frac{1}{k'^2 + m^2}\right)   
   {k'}^{-2i\nu} 
\right] \,.
\nonumber
\end{align} 
Let us denote the expressions in the square brackets as $I(z,\nu)$ and $I(z,-\nu)$ respectively. 
%
%
It is straightforward to put the propagators on a common denominator and show that  
\begin{align} 
I(z,\nu) &=  
- \left(m^2\right)^{-\tfrac{3}{2}+i\nu}  
\Gamma\left(\tfrac{3}{2}-i\nu\right) 
\Gamma\left(-\tfrac{1}{2}+i\nu\right) 
\nonumber\\ 
&=  
- \left(Q_1^2\right)^{-\tfrac{3}{2}+i\nu}  
\left(z \bar z\right)^{-\tfrac{3}{2}+i\nu}  
\Gamma\left(\tfrac{3}{2}-i\nu\right) 
\Gamma\left(-\tfrac{1}{2}+i\nu\right). 
\end{align} 
Next, the $z$ integrals can also be done. We have 
\begin{align} 
\int_0^1 d z\, z\bar z \, \phi(z) \, I(z,\nu) &= 
- \left(Q_1^2\right)^{-\tfrac{3}{2}+i\nu}  
\Gamma\left(\tfrac{3}{2}-i\nu\right) 
\Gamma\left(-\tfrac{1}{2}+i\nu\right) 
\int_0^1 d z\, 6 
\left(z \bar z\right)^{\tfrac{1}{2}+i\nu}  
\nonumber\\ 
&= -6 \sqrt{\pi} \, 2^{-2-2i \nu} 
\left(Q_1^2\right)^{-\tfrac{3}{2}+i\nu}  
\Gamma\left(\tfrac{3}{2}-i\nu\right) 
\Gamma\left(-\tfrac{1}{2}+i\nu\right) 
\frac{ 
\Gamma\left(\tfrac{3}{2}+i\nu\right)}{\Gamma\left(2+i\nu\right)}\,, 
\end{align} 
since the integral over $z$ is just the definition of the Euler beta function. 
Finally this yields the Mellin transform 
\begin{align} 
f_\omega(r^2=0) = & 
\, 9\pi^2 \alpha_s^2 \alpha_{em} \frac{N_c^2-1}{N_c^2} f_\rho^2 Q_1 Q_2 
\int_{-\infty}^{\infty} d\nu\;   
\frac{1}{\omega - \omega(\nu)} 
\left(Q_1^2\right)^{-3/2+i\nu} \left(Q_2^2\right)^{-3/2-i\nu} 
\nonumber\\ 
 \times & 
\frac{ 
\Gamma^2\left(\tfrac{3}{2}-i\nu\right) 
\Gamma^2\left(\tfrac{3}{2}+i\nu\right) 
\Gamma\left(-\tfrac{1}{2}-i\nu\right) 
\Gamma\left(-\tfrac{1}{2}+i\nu\right) 
}{ 
\Gamma\left(2-i\nu\right) 
\Gamma\left(2+i\nu\right) 
} \;,
\end{align} 
which immediately leads to the final result for the amplitude \eq{ampldef} 
\begin{align} 
A(s, t_{min},Q_1, Q_2) =  
is \, 9\pi^2 \alpha_s^2 \alpha_{em}\frac{N_c^2-1}{N_c^2} f_\rho^2 
\, \frac{1}{(Q_1 Q_2)^2}  
\int_{-\infty}^{\infty} d\nu\;   
e^{\omega(\nu) Y}  \left(\frac{Q_1^2}{Q_2^2}\right)^{i\nu}  
\nonumber\\ 
 \times 
\frac{ 
\Gamma^2\left(\tfrac{3}{2}-i\nu\right) 
\Gamma^2\left(\tfrac{3}{2}+i\nu\right) 
\Gamma\left(-\tfrac{1}{2}-i\nu\right) 
\Gamma\left(-\tfrac{1}{2}+i\nu\right) 
}{ 
\Gamma\left(2-i\nu\right) 
\Gamma\left(2+i\nu\right) 
}. 
\label{final} 
\end{align} 
We define $R\equiv Q_1/Q_2$ and write this as 
\begin{align} 
A(s,t_{min},Q_1, Q_2) =  
is \frac{N_c^2-1}{N_c^2} \; \frac{9\pi^2 \alpha_s^2 \alpha_{em} f_\rho^2}{(Q_1 Q_2)^2} J(Y,R)\;, 
\label{final2} 
\end{align} 
with
\begin{align} 
J(Y,R) =  
\int_{-\infty}^{\infty} d\nu\;   
e^{\omega(\nu) Y}  R^{2 i\nu}  
\frac{ 
\Gamma^2\left(\tfrac{3}{2}-i\nu\right) 
\Gamma^2\left(\tfrac{3}{2}+i\nu\right) 
\Gamma\left(-\tfrac{1}{2}-i\nu\right) 
\Gamma\left(-\tfrac{1}{2}+i\nu\right) 
}{ 
\Gamma\left(2-i\nu\right) 
\Gamma\left(2+i\nu\right) 
}. 
\label{Jdef} 
\end{align}


We now want to evaluate the integral $J$. 
We have three possibilities at hand: (i) numerically, (ii) saddle 
point approximation and (iii) sum over residues of poles 
of the integrand. The first two methods are used to obtain BFKL results in the rest of this paper.

\subsection{Born limit}  \label{born}

A very valuable and non-trivial check on the BFKL result as well as the Born result is the 
 fact that in the limit 
 $Y\to 0$ (no evolution) or alternatively $\alpha_s\to 0$ (the gluons in the 
 ladder do not couple to each other) 
 the BFKL amplitude must reduce to the Born level result.  In Appendix \ref{Bornapp} we show, using method (iii) for the evaluation of the integral, that our BFKL result \eq{final} does indeed reduce to the correct Born level 
results derived in Ref.~\cite{PSW}, first for the special case $R=1$ and the DGLAP-like limit $R\gg 1$, and finally for the general case $R\neq 1$.
Note that the calculations presented in the Appendix 
are done in a completely different 
 way than in the Born level calculation in Ref.~\cite{PSW}, so the agreement is very convincing. 
 
The main results from performing the Born limit are presented in Eqs.\ \eq{BornR1}, \eq{Rbig} and \eq{generalJ}.

\subsection{Saddle point approximation of the BFKL amplitude} 
\label{saddlemethod}

For $R=1$ the integrand has a saddle point at $\nu=0$, but the product of $\Gamma$-functions that 
 multiplies the exponential is not very broad. This means the 
 approximation 
 will yield a too large answer. But let us try 
 anyway. 
 The expression multiplying the exponential in the integrand at $\nu=0$ takes the value  
$\pi^3/4$, so for the case $Q_1=Q_2$ we get 
\begin{align} 
J(Y,1) \sim  \frac{\pi^3\sqrt{\pi}}{4}  
\frac{ e^{4\ln 2 \; \as Y}}{\sqrt{14 \as \zeta(3) Y}}, \quad Y\gg 1, 
\end{align} 
so the BFKL amplitude is 
\begin{align} 
A(s,t=t_{min},Q_1=Q_2=Q)
 \sim i s \pi^5\sqrt{\pi} \frac{9 (N_c^2-1)}{4 N_c^2}   
\frac{\alpha_s^2 \alpha_{em} f_\rho^2}{Q^4}  
\frac{e^{4\ln 2 \; \as Y}}{\sqrt{14 \as \zeta(3) Y}}. 
\end{align} 
 
However, we can do better than this. We can keep the general case $Q_1 \neq 
 Q_2$ with $R\equiv Q_1/Q_2$.  The integral is  then
\begin{align} 
J(Y,R) =  
\int_{-\infty}^{\infty} d\nu\;   
e^{\omega(\nu) Y + 2 i\nu \ln R} g\left(\tfrac{1}{2}+ i \nu\right) 
\end{align} 
where $g\left(\gamma=\frac{1}{2}+ i \nu \right)$ was defined in Eq.~\eq{defg}. Expanding the exponent to second order we see that the saddle point is shifted:
\begin{align} 
\omega(\nu) Y + 2 i\nu \ln R  
& \, \sim \, 
\omega(0) Y + 2 i\ln R \; \nu + \frac{\omega''(0)Y}{2} \nu^2  
\nonumber\\   
& \sim \, 
\frac{\omega''(0)Y}{2} \left( \nu + \frac{2i\ln R}{\omega''(0)Y}\right)^2 
+\frac{2\ln^2 R}{\omega''(0)Y} + \omega(0)Y,
\end{align} 
and we can shift the integration variable accordingly to get a Gaussian integral. 
The result is now 
\begin{align} 
J(Y,R) \sim \frac{\pi^3\sqrt{\pi}}{4}  
\frac{ e^{4\ln 2 \; \as Y}}{\sqrt{14 \as \zeta(3) Y}} 
\exp\left(-\frac{\ln^2 R}{14 \as \zeta(3) Y}\right), \quad Y\gg 1, 
\end{align} 
and 
\begin{align} 
A(s,t=t_{min},Q_1,Q_2) \sim i s \, \pi^5\sqrt{\pi} \, \frac{9 (N_c^2-1)}{4 N_c^2}   
\frac{\alpha_s^2 \alpha_{em} f_\rho^2}{Q_1^2 Q_2^2}  
\frac{e^{4\ln 2 \; \as Y}}{\sqrt{14 \as \zeta(3) Y}} 
\exp\left(-\frac{\ln^2 R}{14 \as \zeta(3) Y}\right). 
\label{AsaddleR}
\end{align}

\begin{figure}[tbp]
\begin{center} 
\epsfig{file=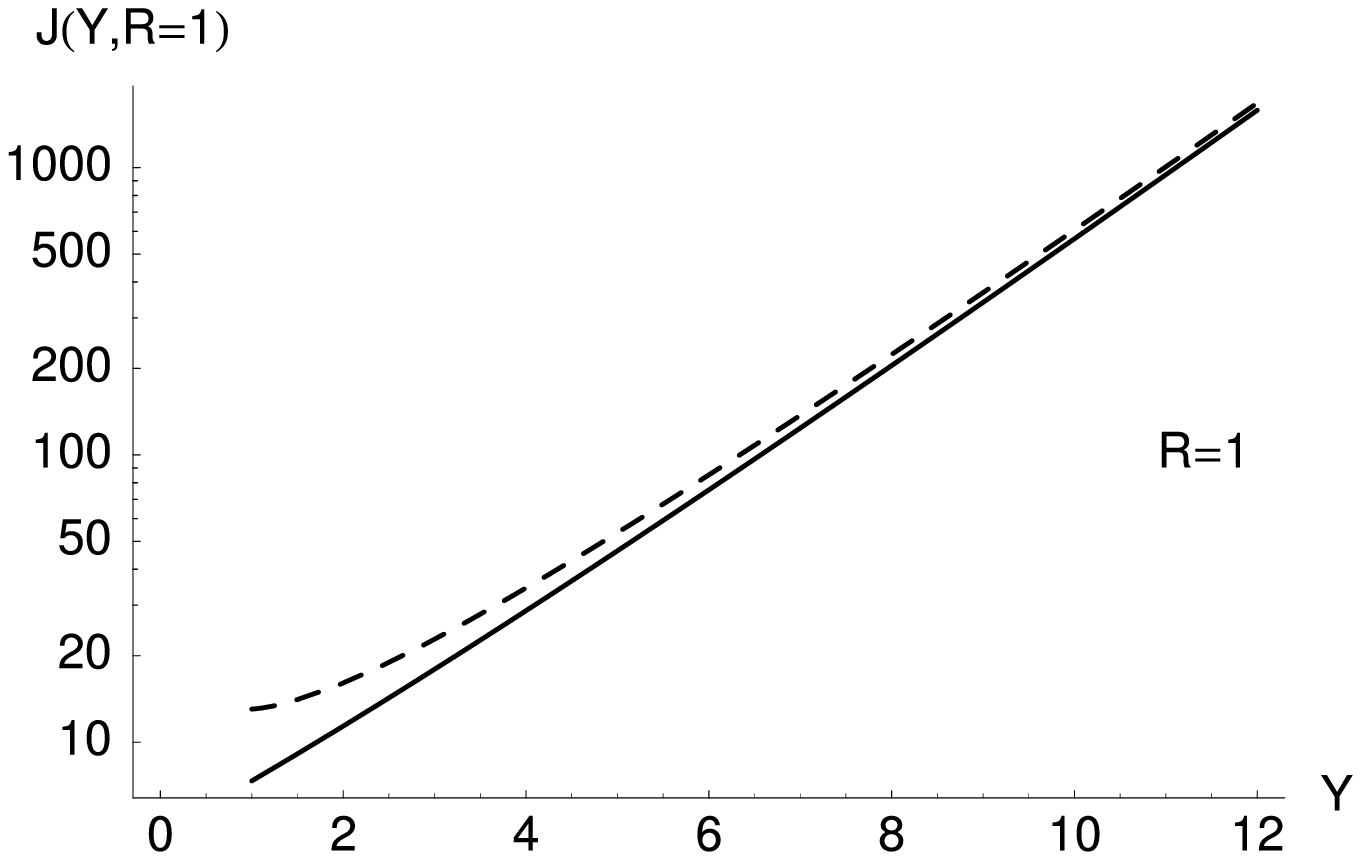,width=0.49\columnwidth} 
\epsfig{file=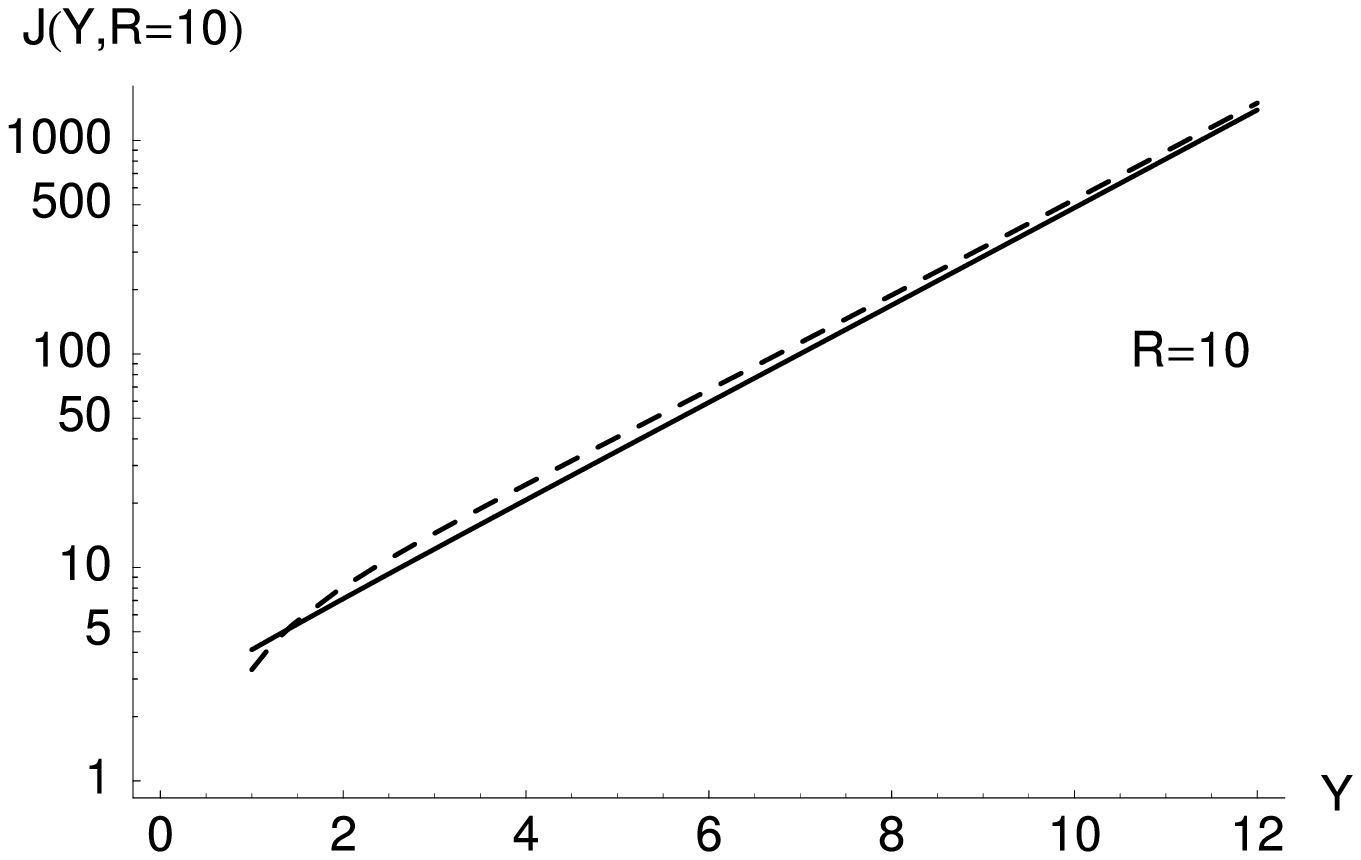,width=0.49\columnwidth} 
\end{center} 
\caption{Exact numerical result (solid line) and saddle point approximation (dashed) to the integral $J(Y,R)$ for $R=1$ (left) and  $R=10$ (right). \label{fig-saddle}} 
\end{figure}

These approximations can be compared to numerical evaluations of the integral
for each value of $Y$. This is illustrated in Fig.\ \ref{fig-saddle}. We see
that the saddle point approximation
 gives the correct asymptotic behavior for large $Y$ and is less accurate for
 smaller $Y.$ For $R=1$ the approximate answer is about 40\% too big for $Y\sim 2$
 and about 10\% too big for $Y\sim 10$. For $R=10$ the approximation is slightly
 better. Note that the Linear Collider is likely to test regions with $Y \gtrsim 5$, where the saddle point approximation works relatively well.

\subsection{Leading order results} \label{results}

To obtain definite predictions we need to define our parameters. The $\rho$ meson decay constant and the electromagnetic coupling take the values $f_\rho=216$ MeV~\cite{frho} and $\alpha_{em}=1/137$, respectively. 
There are furthermore three parameters in the calculation:\ $\alpha_s$ in the prefactor, which gives the strength of the coupling of the pomeron to the impact factor; $\as$ in the BFKL exponent, which gives the strength of the coupling of the gluons inside the pomeron; and the energy scale of the rapidity $Y$.

For all cases with a running strong coupling we use a three-loop running $\alpha_s(\mu^2)$ \cite{Bethke:2000ai} with $\mu^2 = c_\alpha Q_1 Q_2$. Unless otherwise stated, we use this running coupling with $c_\alpha =1$ in the prefactor of the amplitude. 

At LL accuracy, $\as$ is a fixed parameter, i.e., it does not run with the gluon momenta in the BFKL ladder. We choose to, however, let it depend on the given $Q_1$ and $Q_2$, which are external to the pomeron but provide a reasonable choice; we thus choose $\as=\tfrac{N_c}{\pi} \alpha_s(Q_1 Q_2)$. The pomeron intercept is determined by $\as$, and it is known to be too large when comparing to HERA data. Our chosen values give quite large pomeron intercepts, but we do not want to artificially suppress the growth by choosing very small values of $\as$. Instead we will see in  Section \ref{NLLsection} that the growth becomes slower when NLL corrections to the BFKL evolution are included.

The rapidity is defined as 
\begin{equation}
Y = \ln \left(c_Y\frac{s}{Q_1 Q_2}\right),
\end{equation}
where $c_Y$ is a constant that is not constrained at LL accuracy. As discussed in \cite{brodsky} this constant is related to the average attained values of $z_{1,2}$ in the process. The authors of \cite{brodsky} chose a very small value $c_Y=0.01$. We estimate the corresponding effect more conservatively and choose $c_Y=0.3$ for the cross section predictions shown below (see also \cite{Boonekamp}). 

We will now investigate the sensitivity to these choices.
Note that all the results shown in this Section have been obtained by numerical evaluation of the integral over $\nu$ and not by the saddle point approximation.

\begin{figure}[tbp]
\begin{center}
\epsfig{file=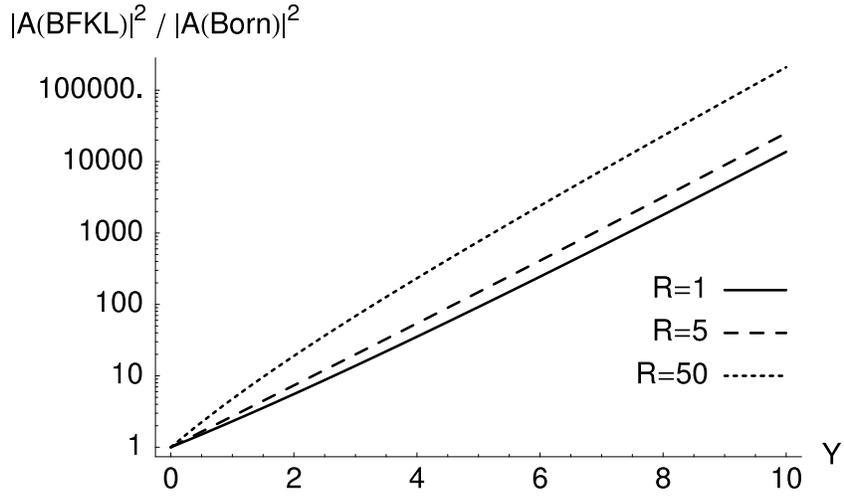,width=\wid}
\end{center}
\caption{Ratio of the leading order BFKL cross section $\left. \frac{d
      \sigma}{dt} \right|_{t_{min}}$ to the Born level cross section as a
  function of the rapidity $Y$, using $\as=0.2$. 
The solid curve is for $R=Q_1/Q_2=1$, the dashed curve is for $R=5$, and the dotted curve is for $R=50$.
\label{fig-comp-y}}
\end{figure}

\begin{figure}[tbp]
\begin{center}
\epsfig{file=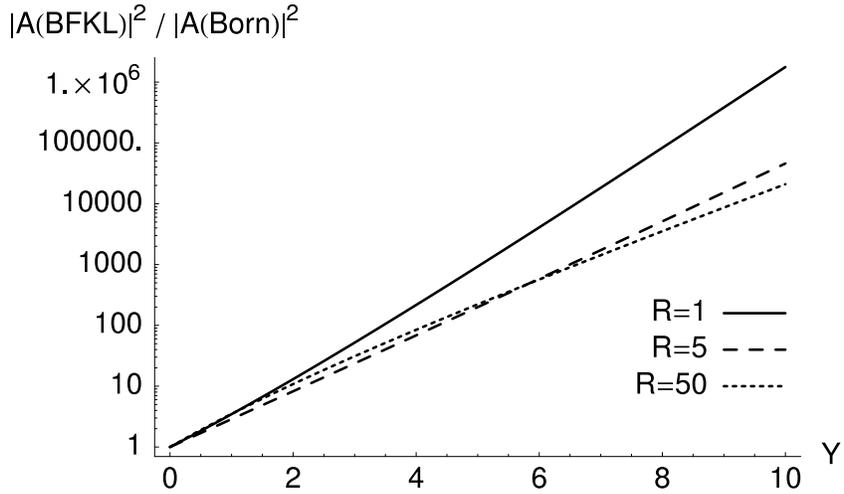,width=\wid}
\end{center}
\caption{Same as Fig.\ \protect\ref{fig-comp-y} but with $\as=\tfrac{N_c}{\pi} \alpha_s(Q_1 Q_2)$, where $Q_2$ is fixed at $Q_2=2$ GeV and $Q_1 = R Q_2$.
\label{fig-comp-y-run}}
\end{figure}

We begin by comparing the energy dependence of the BFKL cross section and the Born cross section. Fig.\ \ref{fig-comp-y} shows the ratio of the differential cross sections $d \sigma/d t|_{t=t_{min}}$ calculated from BFKL and at the Born level, as a function of the rapidity $Y$ for three choices of $Q_1$ and $Q_2$, and for a fixed value $\as=0.2$. There is clearly a strong $Y$-dependence, as seen from \eq{AsaddleR}. Note that all of the $Y$-dependence of this ratio comes from the BFKL amplitude, since the Born level result is independent of the energy. Fig.\ \ref{fig-comp-y-run} shows the same plot but using our ``standard'' choice of $\as=\tfrac{N_c}{\pi} \alpha_s(Q_1 Q_2)$. The difference between Figs.\ \ref{fig-comp-y} and \ref{fig-comp-y-run} serves to illustrate the sensitivity to the choice of $\as$, and shows that using a $Q$-dependent coupling for the pomeron decreases the growth with energy for increasing virtualities.

\begin{figure}[tbp]
\begin{center}
\epsfig{file=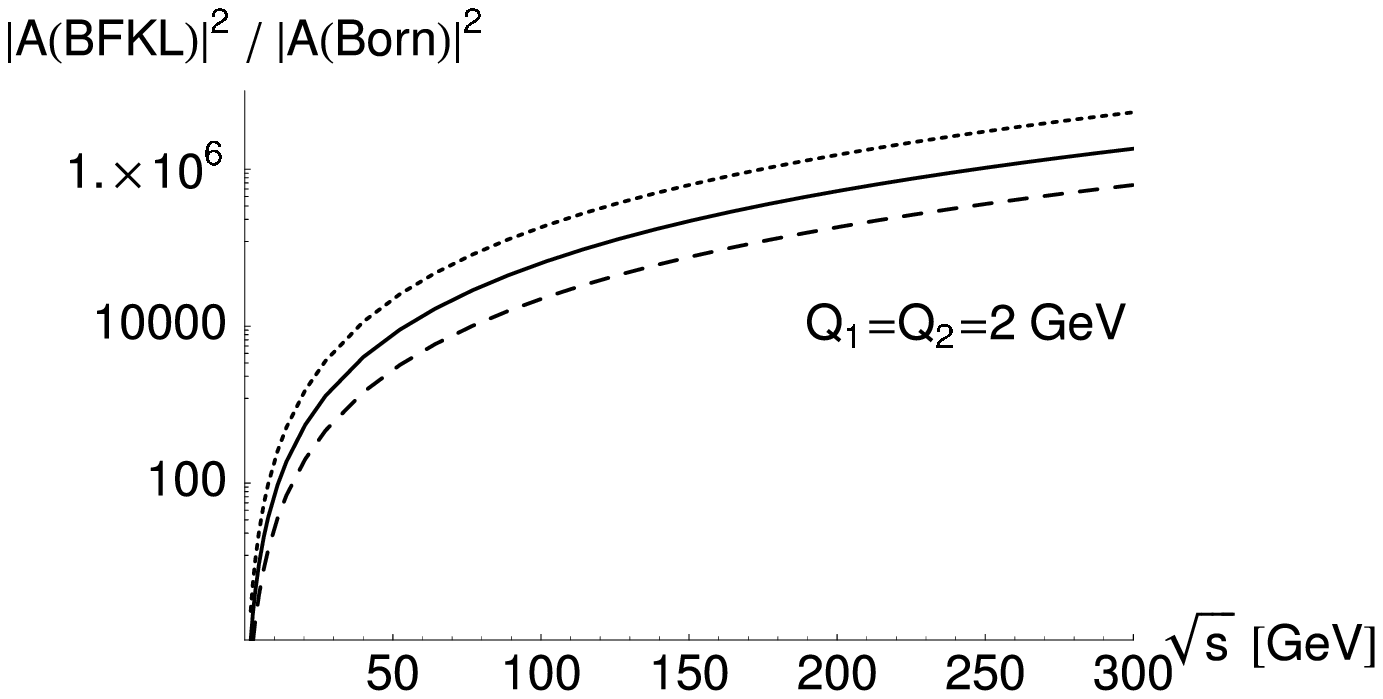,width=0.49\columnwidth} 
\epsfig{file=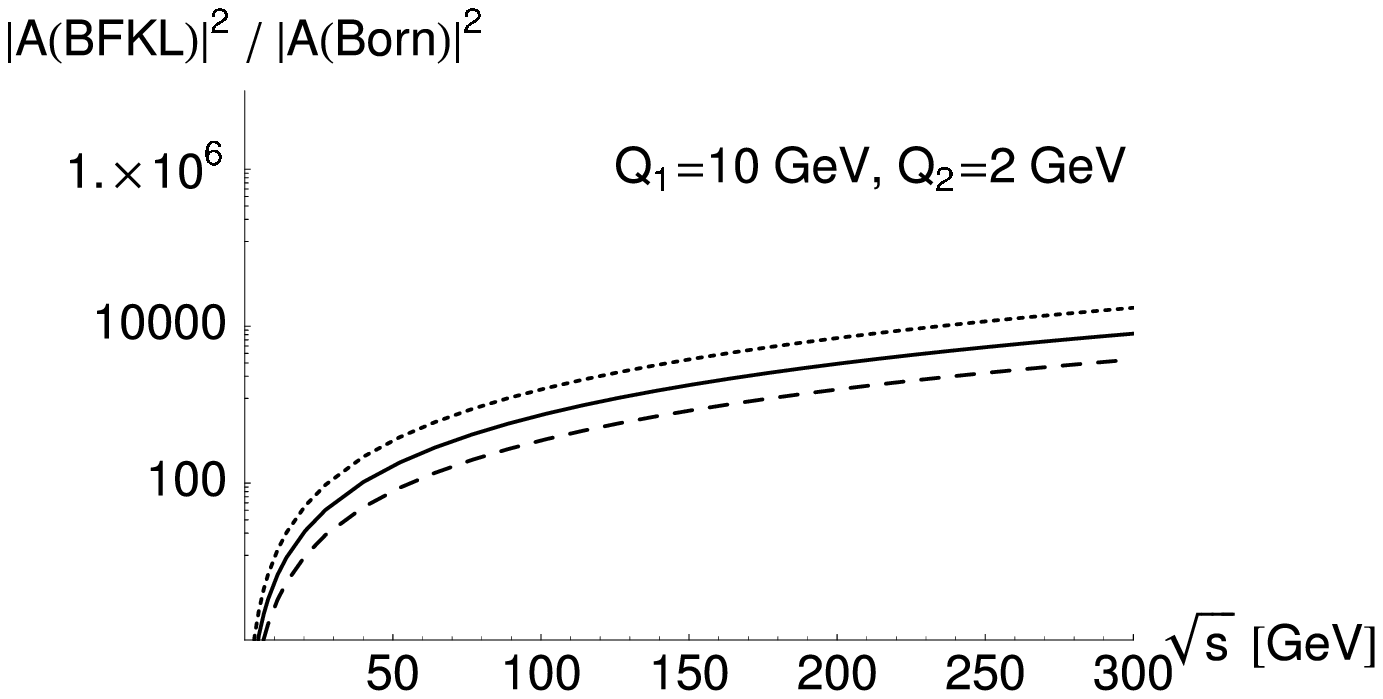,width=0.49\columnwidth} 
\end{center}
\caption{Ratio of the leading order BFKL cross section $\left. \frac{d
      \sigma}{dt} \right|_{t_{min}}$
 to the Born level cross section as a function of the center-of-mass energy
 $\sqrt s$, 
for $Q_1=Q_2=Q=2$ GeV (left) and $Q_1=10$ GeV and $Q_2=2$ GeV (right).
The different curves in each plot correspond to three different definitions of the rapidity variable, $Y=\ln (c_Y s/(Q_1 Q_2)$. The solid curves are for $c_Y=1$, dashed curves are for $c_Y=1/2$, and dotted curves are for $c_Y=2$. The scale of $\as$ is given by $c_\alpha=1$.
\label{fig-comp-s}}

\end{figure}
\begin{figure}[tbp]
\begin{center}
\epsfig{file=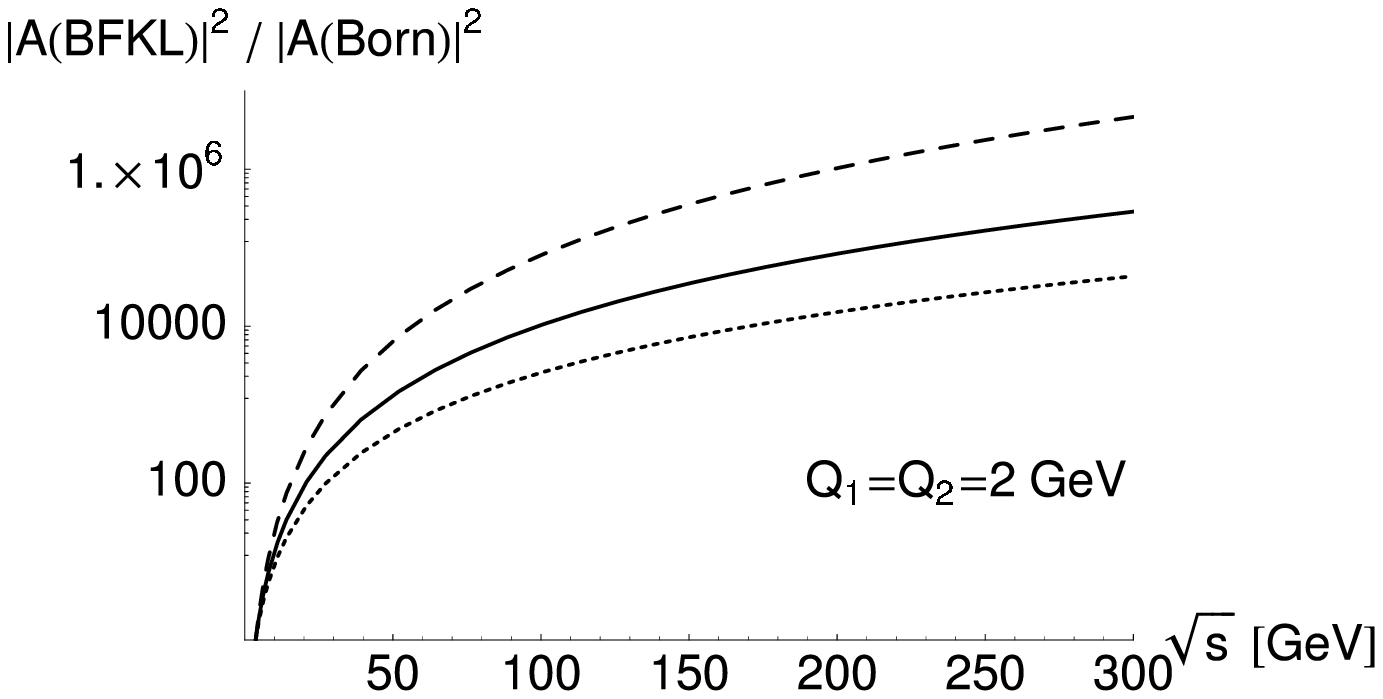,width=0.49\columnwidth} 
\epsfig{file=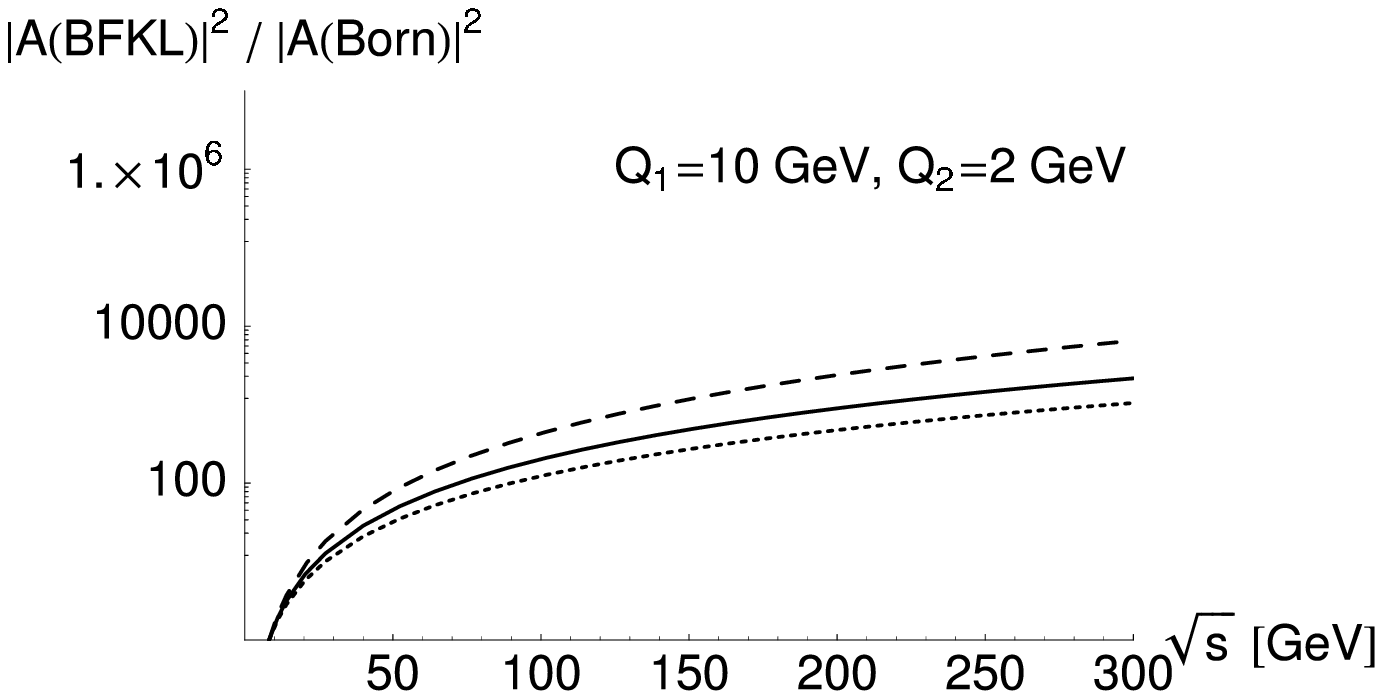,width=0.49\columnwidth} 
\end{center}
\caption{Same as Fig.\ \protect\ref{fig-comp-s}, but the different curves in each plot correspond instead to three different scale choices in $\as$.
The solid curves are for $c_\alpha=1$, dashed curves are for $c_\alpha=1/2$, and dotted curves are for $c_\alpha=2$. The scale of $Y$ is given by the standard $c_Y=0.3$.
\label{fig-comp-s-alpha}}
\end{figure}

In Fig.\ \ref{fig-comp-s} we show the same ratios as a function of the center-of-mass energy $\sqrt s$ for $Q_1=Q_2=2$ GeV in the left plot and for 
$Q_1=10$ GeV, $Q_2=2$ GeV in the right plot. The three different curves represent different choices of the energy scale in the definition of $Y$, corresponding to three different values of the parameter $c_Y$. This freedom to change the scale introduces an additional uncertainty in the results. 
In Fig.\ \ref{fig-comp-s-alpha} we show the same kind of plot, but varying instead the parameter $c_\alpha$ in the argument of $\as$ to highlight the uncertainty coming from the choice of scale in $\alpha_s$. Note that the parameters $c_Y$ and $c_\alpha$ both affect the argument of the BFKL exponential, and thus the energy evolution.

\begin{figure}[tbp]
\begin{center}
\epsfig{file=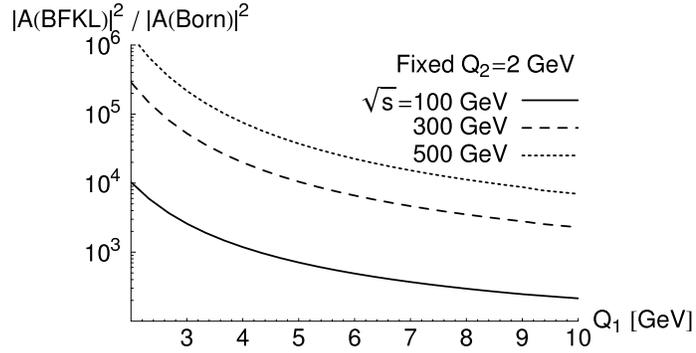,width=\wid}
\end{center}
\caption{Ratio of the leading order BFKL cross section $\left. \frac{d
      \sigma}{dt} \right|_{t_{min}}$
 to the Born level cross section as a function of the virtuality $Q_1$ for fixed $Q_2=2$ GeV, for three different energies $\sqrt s$ and standard scale choices $c_\alpha,c_Y$ as defined in the text.
\label{fig-comp-q}}
\end{figure}

Finally, in Fig.\ \ref{fig-comp-q} we show the ratio as a function of $Q=Q_1=Q_2$. Thus we see that the BFKL prediction differs from the Born level prediction in all kinematical variables, which allows testing BFKL dynamics experimentally, and possibly fitting the free parameters.

We have now seen that there are several uncertainties in the calculated amplitude. This has to be kept in mind when viewing the cross section predictions that will follow.

\begin{figure}[tbp]
\begin{center}
\epsfig{file=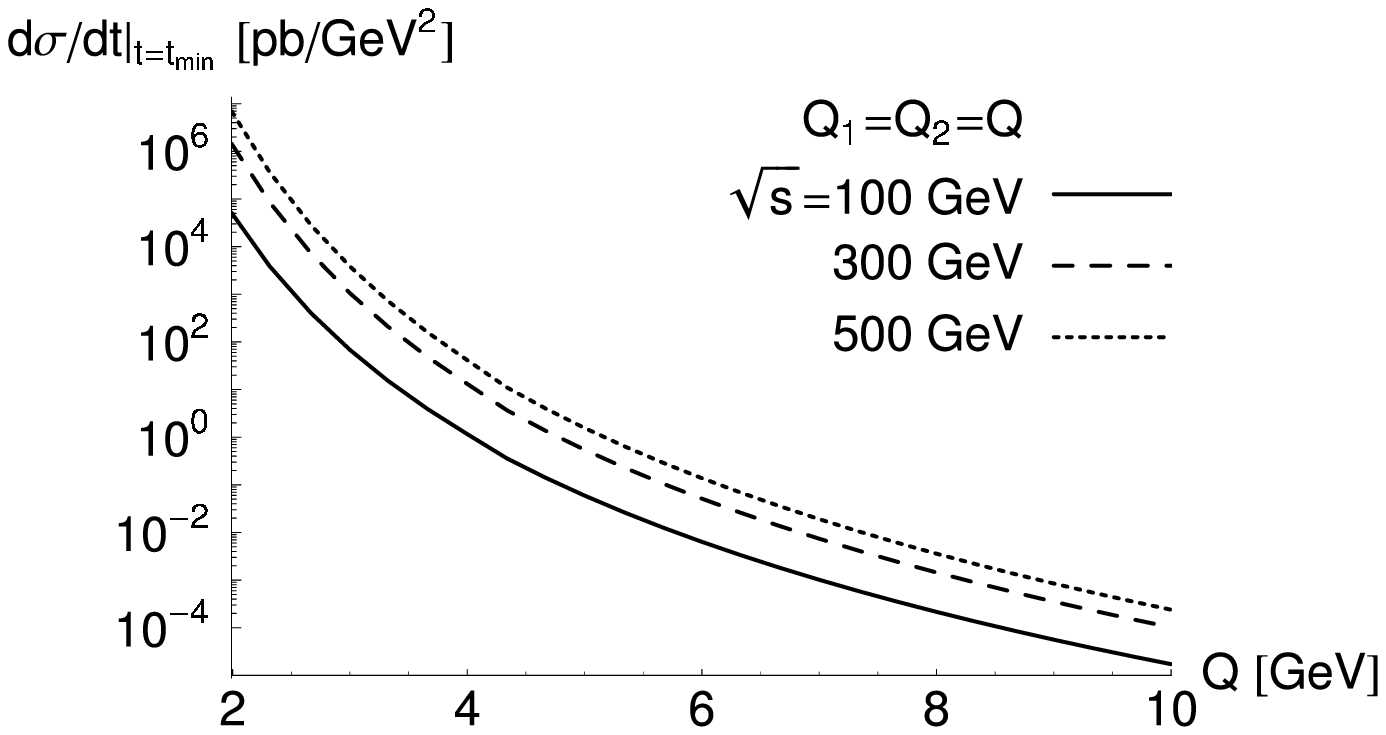,width=\wid}
\end{center}
\caption{Differential cross section $d\sigma/d t|_{t=t_{min}}$ as a function of the photon virtuality $Q=Q_1=Q_2$ for center-of-mass energies
$\sqrt s =$ 100 GeV, 300 GeV and 500 GeV, with standard scale choices as defined in the text.
\label{sigma-q}}
\end{figure}

\begin{figure}[tbp]
\begin{center}
\epsfig{file=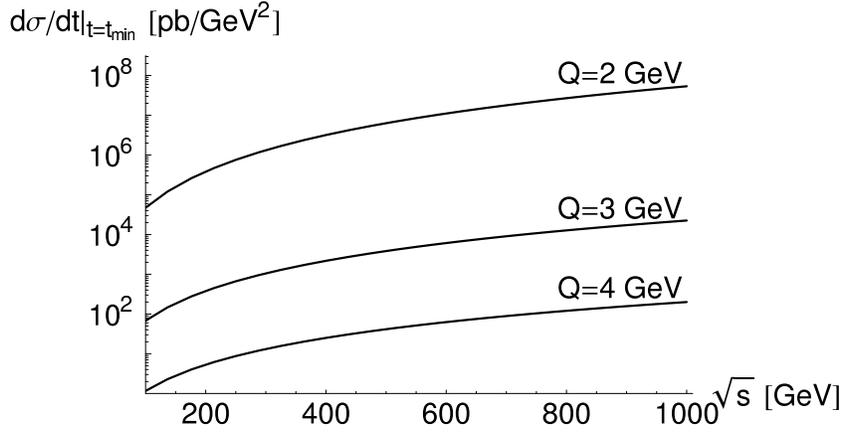,width=\wid}
\end{center}
\caption{Differential cross section $d\sigma/d t|_{t=t_{min}}$ as a function of the photon--photon center-of-mass energy
$\sqrt s$ for photon virtualities $Q=Q_1=Q_2=$ 2 GeV, 3 GeV and 4 GeV.
\label{sigma-s}}
\end{figure}

\begin{figure}[tbp]
\begin{center}
\epsfig{file=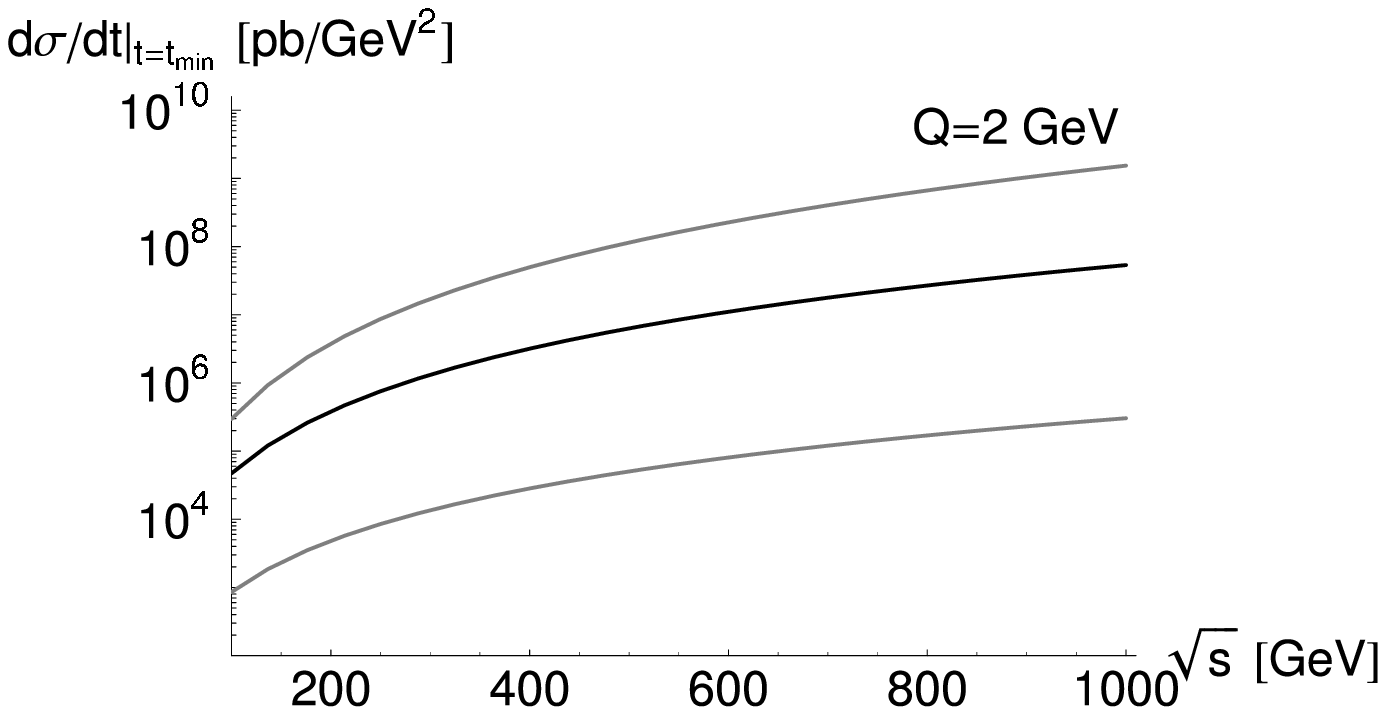,width=\wid}
\end{center}
\caption{Differential cross section $d\sigma/d t|_{t=t_{min}}$ for photon virtuality $Q=Q_1=Q_2= 2$ GeV for standard parameters $c_\alpha,c_Y$ (black curve) and for
new parameters $c'_\alpha=1/2 \, c_\alpha,\; c'_Y=2\, c_Y$ and  
$c'_\alpha=2 \,c_\alpha,\; c'_Y=1/2\, c_Y$ (upper resp.\ lower gray curves). 
\label{sigma-errorband}}
\end{figure}

In Fig.~\ref{sigma-q} we show the differential cross section $d\sigma/d t|_{t=t_{min}}$ as a function of the photon virtuality $Q$ in the symmetric case $R=1$, i.e.\ $Q=Q_1=Q_2$, for three different energies $\sqrt s$,
and in Fig.~\ref{sigma-s} we show the same cross section as a function of  $\sqrt s$ for three different virtualities $Q$. 

These predictions are made with the parameter choices discussed above and contain the corresponding inherent uncertainties. To get some idea of the possible variation in the magnitude of the cross section because of the parameters, we plot in Fig.~\ref{sigma-errorband} the cross section for $Q=2$ GeV for the standard parameter choices, and for two extreme versions, one where we choose new parameters $c'_\alpha=1/2 \, c_\alpha,\; c'_Y=2\, c_Y$ and one with $c'_\alpha=2 \,c_\alpha,\; c'_Y=1/2\, c_Y$. These curves are plotted in gray and should give some indication of the theoretical uncertainty.

\section{Estimation of next to leading order effects}\label{NLLsection}
 
The BFKL kernel is known to NLL accuracy \cite{NLLBFKL}, and the NLL impact factor for our process has recently been computed \cite{NLLIF}. The calculations to obtain the full cross section are difficult, however, and have not yet been performed. It will therefore be interesting to estimate the NLL effects on our calculation. When the cross section is finally computed it may very well be the first complete NLL cross section to be obtained, and it will be very interesting to compare with the estimation presented here.

To estimate higher order effects we thus implement two improvements to the LL BFKL amplitude. First, we use BLM scale fixing \cite{BLM} for the running of the coupling in the prefactor. Second, we use a renormalization group resummed BFKL kernel, as will be explained below.
 
In Ref.\ \cite{brodsky} it is shown that for the BFKL calculation of the total $\gamma^*\gamma^*$ cross section, the BLM procedure for choosing the scale leads to $\mu^2 = c_\alpha Q_1 Q_2$ with $c_\alpha = e^{-5/3}$, and thus to a larger coupling which will increase the cross section. In the process $\gamma^* p \to V p$ at next-to-leading order \cite{Ivanov:2004zv} the correct scale choice was instead found to be $c_\alpha = e^{-1/2}$. The use of the BLM procedure in exclusive processes has been further discussed in \cite{Anikin:2004jb}.

An approximate BLM scale for our process is found by using the NLL impact factors computed in \cite{NLLIF} and neglecting higher order effects in the BFKL kernel. The BLM procedure then, choosing the scale such that the terms proportional to $\beta_0$ vanish, leads to the simple choice $c_\alpha=1$.

The NLL kernel, taken as it is, is larger than the leading order kernel, and for any reasonable value of $\alpha_s$ it is negative; the pomeron intercept becomes less than one for $\alpha_s\gtrsim 0.15$. Furthermore it has two complex conjugate saddle points which can lead to oscillating cross sections. The perturbative expansion of the BFKL kernel is therefore highly unstable and far from converging. See \cite{Salam,CCS,NLLproblems} for a discussion of these problems. 

Different groups have proposed curing the problems of the NLL kernel by certain resummations and methods to stabilize the expansion. In particular some resummation schemes have been proposed~\cite{Salam,CCS,cconstraint,Khoze} which lead to modified BFKL kernels\footnote{The discussion is most conveniently performed using the variable $\gamma=\frac 1 2 +i\nu$ rather than $\nu$.} $\chi(\gamma,\omega)$ depending on both $\gamma$ and $\omega$. This form comes about because of a resummation which removes unphysical double logarithms in the DGLAP and ``anti-DGLAP'' limits where $\gamma$ is close to 0 or 1 \cite{Salam}, and which resums logarithms from the running of the coupling \cite{CCS}. This means that when performing the inverse Mellin transform in Eq.\ \eq{ampldef}, the position of the pole in the $\omega$ plane is determined by the equation 
\begin{equation}
\omega = \as \chi(\gamma,\omega).
\label{defomeganll}
\end{equation}
This equation implicitly defines the function $\omega_{NLL}(\gamma)$ that we need to perform the integral $J(Y,R)$ over $\nu=i(1/2-\gamma)$. We therefore estimate the NLL BFKL result by using a resummed BFKL kernel in the BFKL exponential, but we keep the LL form of the solution and the impact factors.

\begin{figure}[tbp]
\begin{center}
\epsfig{file=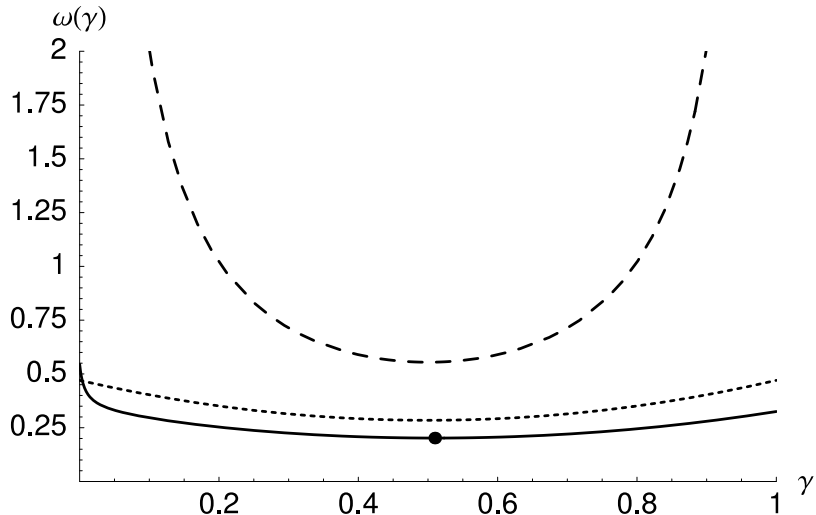,width=\wid}
\end{center}
\caption{The characteristic BFKL functions $\omega_{NLL}(\gamma)$ for the NLL resummed model with $n_f=3$ (solid line) and $n_f=0$ (dotted line) and $\omega(\gamma)$ for LL BFKL (dashed line), for a value of $\as=0.2$. 
 The dot shows the location of the saddle point of $\omega_{NLL}(n_f=3)$,
 $\gamma_s=0.51$, while $\omega(\gamma)$  and $\omega_{NLL}(n_f=0)$ have saddle points in $\gamma=1/2$. %
\label{durham-fig1}}
\end{figure}

The renormalization group (RG) resummed BFKL kernel of Salam \cite{Salam} and Ciafaloni et al.\ \cite{CCS} is actually a resummation of the full NLL kernel, which removes the problems of this kernel. It is however possible to perform such a resummation of the LL kernel, which leads to a result that includes a large part of the corrections coming from the NLL kernel. One such model, based on the general idea of the approach of \cite{Salam,CCS} (more specifically Scheme 4 of \cite{Salam}), was recently proposed by Khoze et al.\ \cite{Khoze}. This approach uses a fixed strong coupling in the BFKL kernel which is essential for the approach here. A running coupling in the BFKL kernel  radically changes the properties of the solutions, and it is no longer possible to evaluate the $\omega$ integral by a simple residue. Therefore we do not pursue it here, although it may be attempted along the lines of \cite{PRSNLL}.

We make one modification to the approach of \cite{Khoze}:\ they use an asymmetric scale choice in the definition of the rapidity (see \cite{Salam,CCS} for a discussion) which was appropriate for their problem under study, but our process is more suited for a symmetric scale choice and we therefore perform the necessary modification. The characteristic function is then expressed as
\begin{equation}
\chi(\gamma,\omega) = \chi_0(\gamma) + \as \chi_1(\gamma,\omega)
\label{durhamchi}
\end{equation}
where $\chi_0(\gamma)\equiv \chi(\gamma)$ is the usual BFKL function \eq{defch}. The correction piece $\chi_1(\gamma,\omega)$ is given by
\begin{equation}
\as \chi_1(\gamma,\omega) =
\frac{1+ \omega A_1(\omega)}{\gamma+\frac{\omega}{2}} -
\frac{1}{\gamma} + 
\frac{1+ \omega A_1(\omega)}{1-\gamma+\frac{\omega}{2}} -
\frac{1}{1-\gamma} -
\omega \chi_0^{\text{ht}}(\gamma), 
\end{equation}
where $\chi_0^{\text{ht}}$ is the higher twist part of $\chi_0$,
\be
\chi_0^{\text{ht}}(\gamma) = \chi_0(\gamma) - \frac{1}{\gamma} - \frac{1}{1-\gamma}
= 2\psi(1) - \psi(1+\gamma) -\psi(2-\gamma).
\ee
$A_1(\omega)$ is obtained from the Mellin transform of the DGLAP splitting function $P_{gg}$ by
\begin{equation}
\frac{1}{2 N_c} P_{gg}(\omega) = \frac{1}{\omega} + A_1(\omega)
\end{equation}
with
\begin{equation}
A_1(\omega) = - \frac{11}{12} - \frac{n_f}{18} + 
\left( \frac{67}{36}-\frac{\pi^2}{6}\right) \omega
+
{\cal O}(\omega^2).
\end{equation}
We will in the following throw away any terms proportional to $\omega$ and only keep singular and constant terms. For $n_f=0$ we then have $A_1(\omega)\simeq -11/12$. It is possible to account for quark loops for $n_f>0$ by replacing \cite{Khoze}
\begin{align}
A_1(\omega) & \to  A_1(\omega) + n_f
\left[ \frac{\as}{4 N_c^2}\frac{1}{\gamma} P_{gq}(\omega) P_{qg}(\omega) -\frac{1}{3} \right]
\nonumber\\ &\simeq
- \frac{11}{12} - \frac{7 n_f}{18} + 
\frac{C_F \as n_f}{6 N_c^2\gamma}
\left( \frac{1}{\omega} -\frac{11}{6}  \right).
\end{align}

The characteristic kernel $\omega_{NLL}(\gamma)$ obtained by solving \eq{defomeganll} with $\chi(\gamma,\omega)$ given by \eq{durhamchi}
is shown both for $n_f=3$ and $n_f=0$ in Fig.~\ref{durham-fig1} together with the LL BFKL kernel. An important feature of the kernel \eq{defomeganll} is that for $n_f=0$ it has no pole at $\gamma=0$. The pole reappears when including the quark loops as shown above. It is also clear that for the resummed kernel both the pomeron intercept and the second derivative at the saddle point are reduced.

To compute the cross section using this model we need to make use of Eq.\ \eq{defomeganll} in performing the integral over $\nu$. This is possible to do purely numerically, by solving the equation \eq{defomeganll} explicitly for each given value of $\gamma$ when performing the integral numerically. However, we may obtain some more insight into the properties of the NLL corrections by instead performing the integral by the saddle point method as in Section \ref{saddlemethod}. We expect the accuracy of the saddle point method to be similar to the LL calculation. For the saddle point calculation we need only compute the position of the saddle point $(\gamma_s,\omega_s)$, where $\omega_s\equiv\omega_{NLL}(\gamma_s)$,
and the particular value  $\omega''_s\equiv\omega''_{NLL}(\gamma_s)$. These can be obtained explicitly in terms of the partial derivatives of the function $\chi(\gamma,\omega)$ as follows.

The chain rule gives
\begin{equation}
\frac{d\omega_{NLL}(\gamma)}{d\gamma}
= \as\left(\frac{\partial\chi(\gamma,\omega)}{\partial\gamma}
+ \frac{\partial\chi(\gamma,\omega)}{\partial\omega}
  \frac{d\omega_{NLL}(\gamma)}{d\gamma}\right)
\end{equation}
so that the saddle point condition gives
\begin{equation}
\left.\frac{d\omega_{NLL}(\gamma)}{d\gamma}\right|_{\gamma=\gamma_s} = 0
\quad\Rightarrow\quad
\left.\frac{\partial\chi(\gamma,\omega_s)}{\partial\gamma}\right|_{\gamma=\gamma_s} = 0.
\end{equation}
We then obtain the second derivative of $\omega_{NLL}(\gamma)$ (using the saddle point condition)
\begin{equation}
\omega''_s=\left.\frac{d^2\omega_{NLL}(\gamma)}{d\gamma^2}\right|_{\gamma=\gamma_s}
= 
\dfrac{\as \left.\frac{\partial^2\chi(\gamma,\omega_s)}{\partial\gamma^2}\right|_{\gamma=\gamma_s}}
{1 - \as
\left.\frac{\partial\chi(\gamma_s,\omega)}{\partial\omega}\right|_{\omega=\omega_s}}.
\end{equation}
Finally, to find the saddle point values $\gamma_s,\omega_s$ we must simultaneously solve the equations
\begin{equation}
\begin{cases}
\displaystyle \left.\frac{\partial\chi(\gamma,\omega_s)}{\partial\gamma}\right|_{\gamma=\gamma_s} = 0,\\
\displaystyle
\omega_s=\as\chi(\gamma_s,\omega_s).
\end{cases}
\end{equation}
The symmetric scale choice with $n_f=0$ gives a kernel symmetric under $\gamma\to 1-\gamma$, so that the saddle point is always located at $\gamma_s=1/2$. This does not hold for an asymmetric kernel obtained either by the asymmetric scale choice, or, as here, by the inclusion of $n_f>0$ effects. The saddle point values $\gamma_s,\omega_s$ must then be determined separately for each value of $\as$.

The residue obtained by the integral over $\omega$ using the $\omega$-dependent kernel is different from the LL case; we have
\begin{equation}
\mathop{\text{Res}}_{\omega=\omega_s} \frac{e^{\omega Y}}{\omega-\as\chi(\gamma_s,\omega)}\
= \frac{e^{\omega_s Y}}{ 1-\dot\omega_s} 
\end{equation}
where we defined the third constant
\begin{equation}\left.
\dot\omega_s\equiv\as\frac{\partial\chi(\gamma_s,\omega)}{\partial\omega}\right|_{\omega=\omega_s}.
\end{equation}
The saddle point evaluation of the integral $J_{NLL}(Y,R)$ now gives
\begin{align} 
A^{NLL}(s,t=t_{min},Q_1,Q_2) \sim
 i s \, \pi^5\sqrt{2\pi} \, \frac{9 (N_c^2-1)}{4 N_c^2}   
\frac{\alpha_s^2 \alpha_{em} f_\rho^2}{Q_1^2 Q_2^2}  
\frac{e^{\omega_s Y}}{\sqrt{\omega''_s Y}} 
\frac{\exp\left(-\frac{2\ln^2 R}{\omega''_s Y}\right)}{1-\dot\omega_s}. 
\label{AsaddleR-NLL}
\end{align}

\begin{figure}[tbp]
\begin{center}
\epsfig{file=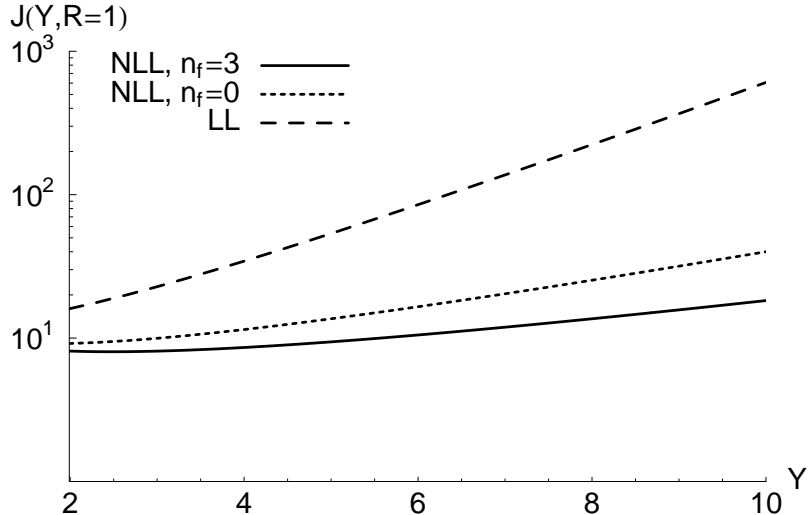,width=\wid}
\end{center}
\caption{Saddle point approximation of the integral $J(Y,R=1)$ for LL BFKL and for the NLL corrected kernel.   
\label{nllintegral}}
\end{figure}

We show the difference between the energy evolution of the LL kernel and the resummed NLL kernel in Fig.\ \ref{nllintegral}. This plot shows the integral $J(Y,R=1)$ for LL and NLL for a fixed value $\as=0.2$. The growth with rapidity is strongly reduced by the NLL effects, and the diffusion pattern is also changed. This is quantified by the pomeron intercept 
$\alpha_{\mathbb P}=\omega_s$, which is reduced from $\alpha_{\mathbb P}=0.55$ to $\alpha_{\mathbb P}=0.20$, and by the second derivative $\omega''_s$ of the kernel, which decreases from $\omega''_s = 28 \as \zeta(3) \simeq 6.73$ at the LL level to $\omega''_s \simeq 1.02$ using the NLL approximation.\footnote{Note that these values of $\omega_s$ and $\omega''_s$ depend on the value of $\as$, so care has to be taken to determine them correctly when using formula (\protect\ref{AsaddleR-NLL}).} The overall normalization is also affected by the factor $1/(1-\dot\omega_s)$, where in this case $\dot\omega_s\simeq-1.51$.

\begin{figure}[tbp]
\begin{center}
\epsfig{file=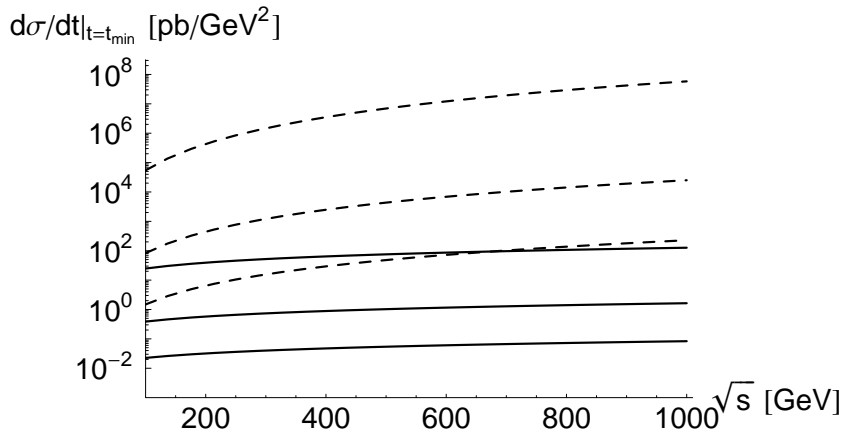,width=\wid}
\end{center}
\caption{Cross section for LL BFKL (dashed lines) and for the NLL corrected kernel (solid lines), using $c_Y=0.3$ and the BLM scale choice $c_\alpha=1$, for the three cases $Q=Q_1=Q_2=$ 2 GeV, 3 GeV and 4 GeV (from top to bottom in the plot).   
\label{nllsigma}}
\end{figure}

The complete cross section prediction from the NLL modified amplitudes with BLM scale choice is shown in Fig.\ \ref{nllsigma} as a function of the energy.  The dashed line shows the LL BFKL result for comparison.

\section{Conclusions}
   
Since we expect the linear collider to be able to cover a quite large region in
rapidity, experiments will allow us to test the dynamics of Pomeron exchange.
The Born approximation estimate \cite{PSW}
of this reaction showed that the process $ \gamma^* \gamma^* \to \rho \rho $ should be measurable by
dedicated experiments at the next linear collider, for virtualities of the photons up to a few GeV$^2$. The increase
of the amplitude that we have obtained when considering BFKL resummation effects is good news for the experimental feasibility of this study.
At leading order the growth with energy is very large, but the next to leading order effects seem to moderate the growth.

This exclusive diffractive reaction may as anticipated become the best tool to investigate the perturbative picture of the hard Pomeron. For example, considering that the full next-to-leading order BFKL calculation of the cross section is likely to appear in the near future \cite{NLLIF}, measurements of this process would be able to test the impact of the NLL corrections and the various approaches proposed in the literature.
Finally, it would also be interesting to pursue this line of research with Odderon exchange processes such as $ \gamma^* \gamma^* \to \pi^0 \pi ^0$ or, through
interference effects in charge asymmetric observables in 
$ \gamma^* \gamma^* \to \pi^+ \pi^- \pi^+ \pi^- $ \cite{HPST}

\paragraph{Note added:} Shortly after submitting this paper, the paper \cite{Ivanov:2005gn} appeared, containing the full NLL calculation alluded to above, and confirming our LL results.
We also note that the BLM scale we obtained under our assumptions in Section \ref{NLLsection} is not changed if the NLL effects on the BFKL kernel are included.

\section*{Acknowledgments}

We thank A.\ Szczurek for discussions on our results. We also thank M.\ Segond for discovering a mistake in our numerical code. 
This  work  is partially supported by the Polish Grant 1 P03B 028 28 
and the French--Polish scientific agreement Polonium.
L.Sz.\ is a Visiting Fellow of 
the Fonds National pour la Recherche Scientifique (Belgium).
R.E.\ would like to thank the Theoretical Physics group at the University of Li\`ege for their hospitality when parts of this work was done.

\appendix

\section{Born limit} \label{Bornapp}

In order to simplify the
 presentation that follows, we will make this check in two steps. 
First, we evaluate the 
 integral $J(0,1)$ exactly, 
 that is, in the case that $R=1$, by picking up all the residues of the integrand in 
 the correct half-plane. 
 Second, the method is generalized to the
 case where $R$ is arbitrary. In the asymptotic
 limit $R\gg 1,$ the leading contribution can then be easily extracted, as was
 also done in \cite{PSW},
and further, an explicit twist expansion
can be written. 
 
 For fixed $R$, the integral is now just a constant number. Changing to the 
 integration variable $\gamma=\tfrac 1 2+i\nu,$ 
we get 
\beq 
\label{Jg}
J(0,R) = 
\frac{1}{R}\int_{1/2-i\infty}^{1/2+i\infty} \frac{d\gamma}{i} g(\gamma) 
\ee 
where 
\beq
\label{defg}
g(\gamma) \equiv 
 R ^{2\gamma}\, g_1(\gamma)
\ee
with
\beq
\label{defg1}
g_1(\gamma) \equiv\frac{\pi^3 \gamma (1-\gamma)}{\Gamma(5/2-\gamma) \Gamma(3/2+\gamma)
  \sin^3(\pi \gamma)}=\frac{f_1(\gamma)}{\sin^3(\pi \gamma)}\,.
\ee
In the following, we will use the auxiliary functions $f$ and $f_1$ defined by
\beq
\label{deff}
g(\gamma) \equiv \frac{f(\gamma)}{\sin^3(\pi \gamma)}
\equiv R^{2\gamma} \frac{f_1(\gamma)}{\sin^3(\pi \gamma)}\,.
\ee
 The function $g(\gamma)$ obviously has triple poles at all integers as given by the $\sin^3 (\pi \gamma)$ factor, except at $\gamma=0,1$ where it has double poles.
 
Let us start with the case $R=1$. The integration contour (in 
 $\gamma$) is parallel to the 
 imaginary axis and crosses the
 real axis between $0$ and $1$. The contour can be closed arbitrarily to the
 right or to the left. We choose to close the contour in the left 
 half-plane and thus to pick up  
all the poles at $\gamma=-n$, $n=0,1,2,\dots$

To extract the triple pole residues at  $\gamma=-n$, $n=1,2,\dots$ 
 we write $\gamma = x-n$ where $n\in \mathbb Z$, $n\neq 0,1$ and $x \in \mathbb R$, $x \ll 1$ so that
\begin{equation}
g(\gamma)=\frac{f(x-n)}{\sin^3 \pi(x-n)} 
\simeq \frac{(-1)^n}{(\pi x)^3} \left[ 1+\frac{\pi^2 x^2}{2}\right] f(x-n)
\end{equation}
The residue at $\gamma=-n$ is the residue at $x=0$, which is by definition for the triple poles
\begin{equation}
\mathop{\text{Res}}_{\gamma=-n}
g(\gamma) 
=
\lim_{x\to 0} \frac{1}{2!} \frac{d^2}{dx^2} \left(x^3
\frac{(-1)^n}{(\pi x)^3} \left[ 1+\frac{\pi^2 x^2}{2}\right] f(x-n) \right).
\end{equation}
Taylor expanding $f(x-n)$ around $x=0$ now easily yields the needed expression
\begin{equation}
\mathop{\text{Res}}_{\gamma=-n} g(\gamma)= \frac{(-1)^n}{2\pi^3}\left( f''(-n)+\pi^2 f(-n)\right).
\label{theresidue}
\end{equation}

This is the first step in the evaluation. We also need a convenient form for $f''(-n)$ that can be summed over $n$. In order to proceed we rewrite $f_1$ without any $\Gamma$-functions as
\begin{equation}
f_1(\gamma) = \frac{\pi^2 \gamma(1-\gamma) \cos \pi \gamma}
{(1/2-\gamma) (1/2+\gamma) (3/2-\gamma)}.
\label{fsimpl}
\end{equation}
From this expression, we get after some 
 algebra the derivatives of $f_1$ up to the second order, at the pole $-n$,
 in the form
\bea
\label{resf1}
f_1(-n) \,& =& \, (-1)^n \,  \frac{\pi^2}{4} \left(\frac{3}{(2 n -1)}+
\frac{2}{(2 n+1)}+ \frac{3}{(2 n +3)} \right),  \\
f'_1(-n) \,& =& \, (-1)^n \,\frac{\pi^2}{2} \left(\frac{3}{(2 n -1)^2}+
\frac{2}{(2 n+1)^2}+ \frac{3}{(2 n +3)^2} \right),  \nonumber \\
f''_1(-n) \,& =& \, -\pi^2\, f_1(-n) +(-1)^n \, 2\pi^2 \left(\frac{3}{(2 n -1)^3}+
\frac{2}{(2 n+1)^3}+ \frac{3}{(2 n +3)^3} \right), \nonumber
\eea
which immediately leads, for $n>0,$ to
\begin{equation}
\mathop{\text{Res}}_{\gamma_1=-n} g_1(\gamma) =
\frac{1}{\pi} \left[
\frac{3}{(2n-1)^3}
+\frac{2}{(2n+1)^3}
+\frac{3}{(2n+3)^3}
\right].
\label{residueg1}
\end{equation}
The sum over all residues in the left half plane is therefore related to the definition of the Riemann $\zeta$ function, so
\begin{equation}
\sum_{n=1}^{\infty} \mathop{\text{Res}}_{\gamma=-n} g_1(\gamma) = \frac{63\zeta(3)-46}{9\pi}.
\end{equation}
To evaluate the integral we should also include the double pole of
$g_1(\gamma)$
in $\gamma=0.$
Defining 
\beq
\label{defh1}
h_1(\gamma)= \gamma^2 \, g_1(\gamma)\,
\ee
one gets 
\beq
\label{resh1}
h_1(0)= \frac{8}{3 \pi}\quad {\rm and } \quad h'_1(0)= -\frac{8}{9 \pi}
\ee
and thus the residue is $-8/(9\pi)$. So finally,
\begin{equation}
\sum_{n=0}^{\infty} \mathop{\text{Res}}_{\gamma=-n} g_1(\gamma) = \frac{7\zeta(3)-6}{\pi}\,.
\end{equation}
For our first goal which was to consider $R=1,$ 
for which $g(\gamma)=g_1(\gamma),$ 
 this sum of residues of $g_1$
is enough 
and thus $J(0,1)=14\zeta(3)-12$. 
The answer for the Born amplitude at $Q_1 = Q_2 = Q$ therefore is 
\begin{align} 
A^{\rm Born}(s,t_{min},Q) = is \frac{N_c^2-1}{N_c^2}\,  \alpha_s^2 \alpha_{em} f_\rho^2 
\, \frac{9\pi^2}{Q^4}  (14\zeta(3)-12) 
\label{BornR1}
\end{align} 
which is exactly the same as in \cite{PSW}. 
 
Let us now study the general case.
 Note first that the limits $R \gg 1$ and $R \ll 1$ are of special physical interest,
since they correspond to the kinematics typical for deep inelastic scattering
on a photon target described through the collinear approximation, {\it i.e.} the usual
parton model \cite{dglap}. From Eqs.~(\ref{Jg}--\ref{defg1}),
the $\gamma$-contour can be closed on the left (resp. right) half-plane for
$R > 1$ (resp. $R < 1$). Let us investigate the case $R> 1$.
As above for $g_1$, the integrand $g$ has infinitely many poles: a double pole at $\gamma=1$ and triple poles at all 
 other integers. Since we are closing the contour in the left half-plane we 
 need only worry about the poles at $\gamma=-|n|$. Using the method that we 
 used above for $R=1$ gives an expression for the residues at theses poles, 
and a related series expansion in powers of $1/R^2$ for $J(0,R).$
It consists of a term corresponding to the double pole at $0$ (which is leading
 at large $R$),
 plus terms that are suppressed with at least a power $1/R^{2n}$ compared to 
 the $\gamma=0$ pole.
 Note that this expansion with respect to $1/R^2,$ for fixed
 values of $Q_2$, corresponds to a twist expansion. The full resummation of
this twist expansion
 can be performed, as we explain now.

Denoting 
\beq
\label{defh}
h(\gamma)=R^{2 \gamma} \, h_1(\gamma)
\ee
one can readily extract the pole of $g(\gamma)$ at 0 through
the relation
\beq
\label{derh}
h'(0)=2 \ln R \, h_1(0) + h'(0)\,,
\ee
which leads to
$$
h'(0)=\frac{16}{3 \pi}\left(\ln R -\frac{1}{6}\right),
$$
and thus to
\begin{align} 
\label{Jpole0}
J(0,R\gg 1) \sim  \frac{32}{3}\left( \frac{\ln R}{R} - \frac{1}{6R}\right), 
\end{align} 
which, inserted in formula \eq{final2} for the amplitude, gives 
\begin{align} 
A^{\rm Born}(s,t_{min},Q_1 =R \, Q_2,Q_2) = is \frac{N_c^2-1}{N_c^2}\,  \alpha_s^2 \alpha_{em} f_\rho^2 
\, \frac{96 \pi^2}{Q_2^4 \, R^2}   
\left( \frac{\ln R}{R} - \frac{1}{6R}\right). \label{Rbig}
\end{align} 
This is again exactly the result as in \cite{PSW}. Let us go further.
Combining Eqs.~\eq{theresidue} and (\ref{defg}--\ref{deff}),
one obtains
\beq
\label{resg}
\mathop{\text{Res}}_{\gamma=-n} g(\gamma)=\mathop{\text{Res}}_{\gamma=-n}
g_1(\gamma) + (-1)^n \frac{2}{\pi^3} \frac{1}{R^{2 n}}(\ln^2 R \, f_1(-n)+
\ln R \, f'_1(-n))\,
\ee
Using the result \eq{resf1} for $f_1$ and  $f'_1,$ 
and relating the obtained series to the ${\rm Li_2}$ and ${\rm Li_3}$ series
\beq
\label{Li2}
\lid(z) \,=\, \sum\limits_{k=0}^\infty \frac{z^k}{k^2},
\qquad
\qquad
\lit(z) \,=\, \sum\limits_{k=0}^\infty \frac{z^k}{k^3},
\ee
one finally gets, after using Eq.~\eq{residueg1} and \eq{Jpole0},
\bea
\label{generalJ}
&&J(0,R)=-6\left(R+\frac{1}{R}\right) -6 \left(R -\frac{1}{R}\right) \ln R -3 \,
\left(R+\frac{1}{R} \right)\, \ln^2 R \\
&& + \left[\frac{3}{R} + 2+ 3 R^2\right] \left[
  \left(\ln \left(1+\frac{1}{R} \right)-\ln \left(1-\frac{1}{R} \right)\right)
  \frac{\ln^2 R}{2} \right. \nonumber \\
&& \left. + \left(\lid \left(\frac{1}{R} \right)-\lit
    \left(-\frac{1}{R} \right)\right) \ln R + \left(\lit \left(\frac{1}{R} \right)-\lit
    \left(-\frac{1}{R} \right)\right)\right],\nonumber 
\eea
which is in agreement with the result obtained in  \cite{PSW}, although the
formula given there is expressed in a different way. After some transformations using the Landen
relations \cite{Landen} for $\lid$ and $\lit$ one finds that the two expressions  match.
The obtained formula \eq{generalJ} is very convenient to obtain the twist
expansion. We illustrate this in the following formula, by giving the first
twist correction to the amplitude,
\bea
\label{amplitudecorrec}
&&\hspace{-.5cm} A^{\rm Born}(s,t_{min},Q_1 =R \, Q_2,Q_2) \\
&&\hspace{-.5cm} = is \frac{N_c^2-1}{N_c^2}\,  \alpha_s^2 \alpha_{em} f_\rho^2 
\, \frac{96 \pi^2}{Q_2^4 \, R^2}   
\left[ \frac{1}{R}\left(\ln R - \frac{1}{6}\right)+\frac{1}{R^3}\left(\frac{2}{5}
  \ln^2 R +\frac{47}{75} \ln R +\frac{1307}{2250} \right)\,+\,\cdots\,\right]. \nonumber 
\eea

 

\end{document}